\documentclass[conf]{new-aiaa}
\usepackage[utf8]{inputenc}

\usepackage{graphicx}
\usepackage{amsmath}
\usepackage[version=4]{mhchem}
\usepackage{siunitx}
\usepackage{longtable,tabularx}
\usepackage{tikz,pgfplots,tikz-3dplot}
\usepackage{pgfplots}
\usepackage{textcomp}
\usetikzlibrary{positioning, plotmarks, arrows, calc, shadings, patterns,
	decorations.pathreplacing, calc,shapes,external}

\pgfplotsset{compat=newest}
\usetikzlibrary{plotmarks}
\usetikzlibrary{arrows.meta}
\usepgfplotslibrary{fillbetween}
\usepgfplotslibrary{patchplots}
\usepackage{grffile}
\usepackage{multirow}
\usepackage{array}

\usepackage{url}

\tikzstyle{blockdiag}	= [node distance=5mm, >=stealth', semithick]
\tikzstyle{block}			= [draw, rectangle, minimum width=1cm, minimum 
height=.8cm]
\tikzstyle{sum} = [draw,circle,inner sep=0pt, minimum size=6pt]
\tikzstyle{connector} = [draw,circle,inner sep=0pt, minimum size=2pt, 
fill=black]
\tikzstyle{branch} = [circle,inner sep=0pt,minimum size=1mm,fill=black,draw=black]
\tikzstyle{gain} = [draw,regular polygon, regular polygon 	sides=3,thick,minimum height=3em,minimum width=4em, rotate=30]
\tikzstyle{bguide} = [rectangle,minimum height=3em,minimum	width=4em]
\tikzstyle{line} = [thick]
\tikzstyle{guide} = [anchor=center]
\tikzstyle{input} = [coordinate]
\tikzstyle{output} = [coordinate, node distance=1cm]

\pgfdeclarelayer{background}
\pgfdeclarelayer{foreground}
\pgfsetlayers{background,main,foreground}

\definecolor{mycolor1}{rgb}{0.00000,0.44700,0.74100}%
\definecolor{mycolor2}{rgb}{0.85000,0.32500,0.09800}%
\definecolor{mycolor3}{rgb}{0.92900,0.69400,0.12500}%
\definecolor{mycolor4}{rgb}{0.49400,0.18400,0.55600}%
\definecolor{mycolor5}{rgb}{0.46600,0.67400,0.18800}%

\definecolor{mycolor6}{rgb}{0.00000,0.05263,0.97368}%
\definecolor{mycolor7}{rgb}{0.00000,0.10526,0.94737}%
\definecolor{mycolor8}{rgb}{0.00000,0.15789,0.92105}%
\definecolor{mycolor9}{rgb}{0.00000,0.21053,0.89474}%
\definecolor{mycolor10}{rgb}{0.00000,0.26316,0.86842}%
\definecolor{mycolor11}{rgb}{0.00000,0.31579,0.84211}%
\definecolor{mycolor12}{rgb}{0.00000,0.36842,0.81579}%
\definecolor{mycolor13}{rgb}{0.00000,0.42105,0.78947}%
\definecolor{mycolor14}{rgb}{0.00000,0.47368,0.76316}%
\definecolor{mycolor15}{rgb}{0.00000,0.52632,0.73684}%
\definecolor{mycolor16}{rgb}{0.00000,0.57895,0.71053}%
\definecolor{mycolor17}{rgb}{0.00000,0.63158,0.68421}%
\definecolor{mycolor18}{rgb}{0.00000,0.68421,0.65789}%
\definecolor{mycolor19}{rgb}{0.00000,0.73684,0.63158}%
\definecolor{mycolor20}{rgb}{0.00000,0.78947,0.60526}%
\definecolor{mycolor21}{rgb}{0.00000,0.84211,0.57895}%
\definecolor{mycolor22}{rgb}{0.00000,0.89474,0.55263}%
\definecolor{mycolor23}{rgb}{0.00000,0.94737,0.52632}%
\definecolor{mycolor24}{rgb}{0.00000,1.00000,0.50000}%

\newcommand{\sys}[1]{{#1}} 

\newtheorem{mytheo}{Theorem}

\newcommand{\norm}[1]{\left\|#1\right\|}
\newcommand{\abs}[1]{\left|#1\right|}
\newcommand{\field}[1]{\mathbb{#1}}
\newcommand{\R}{\field{R}}

\newcommand{\bmtx}{\begin{bmatrix}}
	\newcommand{\emtx}{\end{bmatrix}}
\newcommand{\bsmtx}{\left[ \begin{smallmatrix}} 
	\newcommand{\esmtx}{\end{smallmatrix} \right]} 
\newcommand{\bmatarray}[1]{\left[\begin{array}{#1}}
	\newcommand{\ematarray}{\end{array}\right]} 

\graphicspath{{figures/}}

\title{Robust Mode Transition for Spacecraft Attitude Control}

\author{Emily Burgin\footnote{Research Assistant, Chair of Flight Mechanics and Control, emily.burgin@tu-dresden.de.}, Carl-Johann Winkler \footnote{Research Assistant, Chair of Flight Mechanics and Control, carl-johann.winkler@tu-dresden.de.}, Felix Biert\"umpfel \footnote{Postdoctoral Associate, Chair of Flight Mechanics and Control, felix.biertuempfel@tu-dresden.de} and Harald Pfifer \footnote{Professor, Chair of Flight Mechanics and Control, harald.pfifer@tu-dresden.de}}
\affil{Technische Universit\"at Dresden, 01307 Dresden, Germany}

\author{Pedro Simplicio \footnote{GNC Engineer, Guidance, Navigation \& Control Section, pedro.simplicio@ext.esa.int}}
\affil{Aurora Technology for ESA-ESTEC, 2201 AZ Noordwijk, The Netherlands}

\begin{document}
	
	\maketitle
	
	\begin{abstract}
		This paper proposes the design of a single linear parameter-varying (LPV) controller for the combined control and smooth transition between two modes in a spacecraft mission. Current industry practice for transitioning between different controller modes is to use a discrete switching approach. When predefined criteria are satisfied, the controller of one mode is turned off and the controller of the other is initialised, resulting in an undesirable transient behaviour. In addition, each controller must individually undergo a rigorous verification and validation (V\&V) process. A single controller synthesised using LPV methods streamlines the V\&V process and improves the transient behaviour. The proposed design follows a mixed-sensitivity control scheme with LPV weights that are derived from the performance and robustness requirements of the individual modes. The controller is synthesised by minimising the induced $\mathcal{L}_2$-norm of the closed-loop interconnections between the controller and weighted plant. The performance and robustness of the controller is demonstrated on an acquisition and pointing task of a flexible satellite through a Monte-Carlo campaign.
	\end{abstract}
	
%
	\section{Introduction} \label{sec:intro}
\lettrine{S}{pacecraft} control systems are comprised of multiple modes, each with specifically designed controllers for their individual objectives, as in \cite{espeillac2017gnc,barre2008smos}. Consider a satellite reorientation manoeuvre; a fast initial turning rate is realised with an acquisition controller then command is handed over to a pointing controller which is subject to tighter tracking requirements. To execute this manoeuvre, current industry practice would be to use a discrete switching approach, i.e., when predefined criteria are met the acquisition controller is halted and the pointing controller initialises. This initialisation period causes a discontinuity, leading to undesirable, possibly disruptive transient behaviour in the response \cite{lurie2000single}. Moreover, it cannot be guaranteed that the pointing controller is able to keep the initial transients within attitude or actuation requirements \cite{wie1985roll}. 

The present paper, therefore, proposes combining the two separate mode controllers into a single linear parameter-varying (LPV) controller. This approach greatly streamlines the rigorous verification and validation (V\&V) process \cite{PACE2023647}, as only one algorithm must be verified, where it previously would have been two. Moreover, the controller gains transition from acquisition to pointing in a continuous manner, resolving the need for re-initialisation and as such improving the response performance. The LPV design extends from the principles of $\mathcal{H}_{\infty}$ design, the theory of which is well understood and has extensive literature. The controller is synthesised by minimising the induced $\mathcal{L}_2$-norm of the closed-loop interconnections of the controller and an LPV generalised plant, established using a modified version of a recently proposed mixed-sensitivity weighting scheme developed by Theis et al.~\cite{theisRobustModalDamping2020}. This scheme uses a minimum number of physically interpretable weights that correspond directly to the performance and robustness requirements of the closed-loop system. It has been successfully applied to many use-cases in aerospace \cite{theisRobustModalDamping2020,theisRobustAutopilotDesign2018,biertumpfelRobustLTV}. This scheme has previously been applied to spacecraft attitude control in \cite{burgin2023linear}, demonstrating that satellite pointing control, with stringent requirements, can easily be translated into the mixed-sensitivity framework. Thus, it logically follows that the framework can be extended to use LPV weights in order to synthesise a controller for a problem with changing requirements. In this case, a mode transition attitude control task.

Previous attempts at tackling the smooth mode transition objective were successful in improving the transient response performance. For example, the second mode controller can be initialised with the states of the previous one, although, this requires an intermediate state settling \cite{Geshnizjani2013Thesis}. Alternatively, both controllers can be run simultaneously and their output commands blended together along a pre-defined trajectory, but this increases computational effort. Other works have derived parameter-dependent control laws analytically, either by blending two existing mode controllers \cite{biannic2010lpv,rufus2002real}, or via gain scheduling \cite{kasai2015lmi}. While these methods produce a better time response, they lack statements on robustness and stability during the transition phase -- a clear indicator for the superiority of an LPV representation. The induced $\mathcal{L}_2$-norm provides performance bounds across the entire domain, including the transition period. Thus, robustness and stability are guaranteed while the controller is transitioning between the modes. Although commonly misinterpreted, the same result cannot be said of gain scheduling as each controller is synthesised independently and there are no guarantees for the performance when interpolating between controllers. For LPV synthesis, even when synthesising along a grid, it still considers a continuous domain with respect to the scheduling parameter $\rho$. So, in comparison to gain scheduling, when interpolating an LPV controller between synthesis grid points (with respect to the trajectory of $\rho$ defined in the synthesis), the $\mathcal{L}_2$-norm still holds and the robustness and performance is guaranteed.

The single LPV controller approach is validated in this work on a transition between an acquisition and fine-pointing mode of a flexible spacecraft. The performance is measured by the time it takes for the attitude of the spacecraft to converge to within the pointing mode requirements. Actuator saturation limits must also be respected. The performance and robustness are demonstrated through a Monte-Carlo campaign and analytical robustness margin calculations.

\section{BACKGROUND}
\subsection{ Induced $\mathcal{L}_2$-norm Controller Synthesis for Linear Parameter-Varying Systems}
LPV systems are a type of systems whose state-space matrices depend continuously on a time-varying parameter vector $\rho: \R \rightarrow \mathcal{P}$, where $\mathcal{P} \in \R^{n_\rho}$ is a compact subset chosen based on physical considerations. In addition, the parameter rates of variation $\dot{\rho}$ are assumed to lie within a hyper-rectangle $\dot{\mathcal{P}}$ defined by $\dot{\mathcal{P}}= \{\dot{\rho}(t) \in \R^{n_\rho} | \abs{\rho_i(t)} \leq \nu_i, i=1,\hdots,n_\rho   \}$. Hence, the set of all admissible trajectories is $\mathcal{T} = \{ \rho: \R \rightarrow \mathcal{P} | \, \rho \in \mathcal{C}^1, \rho(t) \in \mathcal{P} \textrm{ and } \dot{\rho}(t) \in \dot{\mathcal{P}} \, \forall t \geq 0\}$. 

An LPV system $P_\rho$ can be represented in the state-space formulation where each matrix is a function of the parameter vector, i.e., $A: \mathcal{P} \rightarrow \R^{n_x \times n_x}$, $B: \mathcal{P} \rightarrow \R^{n_x \times n_u}$, $C: \mathcal{P} \rightarrow \R^{n_y \times n_x}$, and $D: \mathcal{P} \rightarrow \R^{n_y \times n_u}$. An $n_x^{\textrm{th}}$-order LPV system $P_\rho$ is defined by 
\begin{equation}\label{eqn:lpvsystem}
	P_\rho:	\bmtx \dot{x}(t) \\ y(t) \emtx = \bmtx  A(\rho(t)) & B(\rho(t)) \\ C(\rho(t)) & D(\rho(t)) \emtx \bmtx x(t) \\ u(t) \emtx,
\end{equation}
where $x(t) \in \R^{n_x}$ is the state vector, $u(t) \in \R^{n_u}$ the input vector, and $y(t) \in \R^{n_y}$ the output vector. The dependence on $t$ is often omitted from the notation for clarity. 

The performance of an LPV system can be specified in terms of its induced $\mathcal{L}_2$-norm
\begin{equation}\label{eqn:inducedL2norm}
	\norm{P_\rho} = \sup_{u \in \field{L}_2 \setminus \{0\}, \rho \in \mathcal{T}, x(0)=0} \frac{\norm{y}_2}{\norm{u}_2}.
\end{equation}
A generalisation of the Bounded Real Lemma \cite{wu1996induced} provides a sufficient condition to upper bound the norm of a system $\norm{P_\rho}$.

\begin{mytheo}\label{theo:genBRL}
	\cite{wu1996induced}:	 $P_\rho$ is exponentially stable and $\norm{P_\rho} \leq \gamma$ if there exists a continuously differentiable symmetric matrix function $X: \mathcal{P} \rightarrow \R^{n_x \times n_x}$ such that $X(p) \geq 0$ and
	\begin{equation} \label{eq:brl}
		\bmtx XA+A^TX + \partial X & XB \\ B^TX & -I \emtx + \frac{1}{\gamma^2} \bmtx C^T \\ D^T \emtx \bmtx C & D \emtx \leq 0
	\end{equation}
	hold for all $p \in \mathcal{P}$ and $q \in \dot{\mathcal{P}}$, where $\partial X$ is defined as $\partial X(p,q) = \sum_{i=1}^{n_\rho} \frac{\partial X}{\partial \rho_i}(p) q_i$. In (\ref{eq:brl}), the dependence of the matrices on $p$ and $q$ has been omitted to shorten the notation.
\end{mytheo}

This theorem extends to the induced $\mathcal{L}_2$-norm controller synthesis in \cite{wu1996induced}. Consider an open-loop LPV system $G_\rho$ with the state-space formulation as in (\ref{eqn:lpvsystem}) with inputs denoted $[w^T, u^T]^T$ and outputs $[z^T, y^T]^T$, where $w$ and $z$ are measures of performance. The objective is to synthesise a controller $K_\rho$,
\begin{equation}\label{eqn:K}
	K_\rho:	\bmtx \dot{x}_K \\ u \emtx = \bmtx A_K(\rho) & B_K(\rho) \\ C_K(\rho) & D_K(\rho) \emtx \bmtx x_K \\ y \emtx,
\end{equation}
such that the induced $L_2$-gain of the closed-loop interconnection of $G_\rho$ and $K_\rho$, denoted by the lower fractional transformation $F_l(G_\rho,K_\rho)$, is minimised.
\begin{equation}\label{eqn:minimisation}
	\min_{K_\rho} \norm{F_l(G_\rho,K_\rho)}.
\end{equation}
Thus, the optimisation of the performance of the closed-loop system can be solved via parametrised LMI conditions; see \cite{wu1996induced} for details. This synthesis problem involves an infinite collection of LMI constraints parametrised by $(p,q) \in \mathcal{P} \times \dot{\mathcal{P}}$. A remedy to this infinite dimensionality is to approximate the constraints with finite-dimensional LMIs evaluated on a gridded domain of $p$ and $q$. Tools to solve the synthesis problem are readily available; LPVTools \cite{hjartarson2015lpvtools} is used in this paper.

\subsection{Linear Parameter-Varying Mixed-Sensitivity Design Architecture}\label{sec:LPVmixedsens}
\begin{figure}[h!] 
	\centering\begin{tikzpicture}[blockdiag, auto]
	
	\node[block, minimum width=1.2cm] (Plant) {$\sys{P}$};
	\node[sum, left=of Plant] (SumP) {};
	\node[above=of Plant, yshift=-.75cm, minimum width=1.2cm](Pd) {};
	\node[branch, left=of SumP,xshift=-8.5mm] (Branch1) {};
	\node[block, above=of Branch1,yshift=0.25cm] (Wu) {$\!\sys{W_{\!u,\rho}}\!$};
	
	\node[block, dashed,left=of Branch1, xshift=.0cm, minimum width=1.2cm] (Controller) {$\sys{K_\rho}$};
	\node[branch, left=of Controller, xshift=0cm] (BranchE) {};
	\node[sum, left=of BranchE, xshift=-0.72cm] (SumE) {};
	\node[block, above=of BranchE,yshift=0.25cm] (We) {$\!\sys{W_{\!e,\rho}}R_{eu,\rho}^{-1}\!$};
	\node[block, above=of BranchE,yshift=0.25cm, xshift=-14mm] (W1) {$\!\!R_{eu,\rho}\!\!$};
	
	\node[block, above=of SumP,yshift=0.21cm] (Wd) {$\!\!R_{du,\rho}\!\!$};
	%
	%
	
	\draw[<-] (Plant) -- (SumP);
	\draw[<-] (SumE) -- (W1)node[pos = 0.5, left]{$r$};
	\draw[<-] (W1.north) -- +(-0,0.25cm) node[above]{$w_1$};
	\draw[<-] (Wd.north) -- +(0,0.25cm) node[above]{$w_2$};
	\draw[->] (We.north) -- +(0, +.25cm) node[above]{$z_1$};
	\draw[->] (Wu.north) -- +(0, +.25cm) node[above]{$z_2$};
	
	\draw[-] (SumE) -- (BranchE);
	\draw[->] (BranchE) -- (Controller) node[pos=0.5] {$e$};
	\draw[->] (BranchE) -- (We);
	\draw[->] (Controller) -- (SumP) node[pos=0.1] {$u$};
	\draw[->] (Branch1) -- (Wu);
	\draw[->] (Wd) -- (SumP)node[pos = 0.5, left]{$d$};

	\draw[->] ($(Plant.east)$) -| +(+.5cm,-1.0cm) -| (SumE.south) node[pos=0.95,swap] {$-$};
	%
	
	\end{tikzpicture} 
	\caption{LPV weighted four-block mixed-sensitivity problem.}
	\label{fig:mixedSens}
\end{figure}
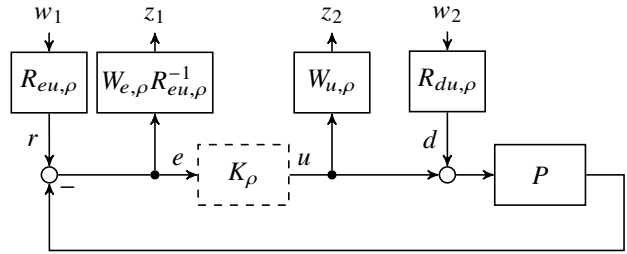

It is common practice to design induced $\mathcal{L}_2$-norm optimal controllers by mixed-sensitivity loopshaping; see, e.g., \cite{ZhouEtAl_95_Optimal, Skogestad2005}. Thus, the theory extends to optimal LPV controllers \cite{wu1996induced}, as demonstrated in \cite{burgin2023linear}. Consider the closed-loop feedback system between a plant $P$ and controller $K_\rho$ (\ref{eqn:K}). Desired closed-loop behaviour can be enforced by minimising the induced $\mathcal{L}_2$-norm of the interconnection between the controller and a weighted, generalised plant $G_\rho$ (\ref{eqn:lpvsystem}) constructed from $P$ and some weights. The weights are responsible for defining the additional performance in/outputs $w$ and $z$. The proposed weighting scheme applied in this paper uses a minimal number of physically interpretable LPV weights that are derived from the robustness and performance requirements of the closed-loop with respect to the scheduling parameter $\rho$. The scheme is shown in Fig.~\ref{fig:mixedSens}. It is adapted from a recently proposed scheme presented for an LTI case in \cite{theisRobustModalDamping2020}. Note that the subscript $\rho$ indicates a system's dependence on the scheduling parameter $\rho$. Throughout this paper, the plant $P$ is assumed to be linear time-invariant (LTI) as only the performance weights change during the mode transition. The theory presented in this paper also extends to LPV synthesis when the plant is parameter-varying, i.e., $P_\rho$.

Defining the output sensitivity function $S_\rho = (I+P K_\rho)^{-1}$, the generalised closed-loop $F_l(G_\rho,K_\rho)$ of the weighted mixed-sensitivity problem is then
\begin{equation} \label{eqn:mixedSens}
		\bmtx z_1 \\ z_2 \emtx = \bmtx W_{e,\rho} R_{eu,\rho}^{-1} & 0  \\ 0 & W_{u,\rho} \emtx\bmtx S_\rho & -S_\rho P \\ K_\rho S_\rho & -K_\rho S_\rho P \emtx \bmtx R_{eu,\rho} & 0 \\ 0 & R_{du,\rho} \emtx \bmtx w_1 \\ w_2 \emtx
\end{equation}
where $W_{e,\rho}$ and $W_{u,\rho}$ denote dynamic parameter-varying weights and $R_{eu,\rho}$, and $R_{du,\rho}$ parameter-dependent scaling factors. The central block is referred to as the \textit{four-block problem} and defines four unique closed-loop mappings that are shaped by the designed weights. These four blocks fully describe the performance and robustness of the controlled system. A high magnitude in $W_{e,\rho}$ reduces $S_\rho$ leading to better tracking and disturbance rejection capabilities. A high magnitude in $W_{u,\rho}$ reduces the control effort $K_\rho S_\rho$. Hence, $W_{u,\rho}$ can enforce controller roll-off at high frequencies, e.g., to avoid excitation of flexible modes in a system. The scaling factors are used as the main tuning knobs and are mutually dependent. $R_{eu}$ tunes the desired relationship between command signal $u$ and error $e$. A good initial value is the ratio of allowable pointing error to maximum actuator command, implying that the synthesised controller will command its maximum capacity when the pointing error is about to be violated. Similarly $R_{du}$ defines the relationship between expected disturbance $d$ and actuator response $u$. As a result, a third relationship must be considered $R_{de} = R_{du}R_{eu}^{-1}$ which tunes the error $e$ relative to the expected disturbance; in other words, the disturbance rejection performance.

\section{LPV Control for Satellite Mode Transition}\label{sec:LPVTrans}

This paper applies LPV controller synthesis to combine the traditionally separate modes of a spacecraft attitude reorientation controller, as described in Section \ref{sec:intro}. The controller generates torque commands $\tau_{cmd}$ about the body-frame to establish tracking of the satellite attitude $\theta$ to the guidance (GUI) reference signal $\theta_r$, respecting the actuators’ limitations. It is imperative to recognise the differing requirements and desired response characteristics for the target acquisition phase (denoted $a$)compared to the pointing phase (denoted $p$). A mixed-sensitivity weighting scheme can be easily formulated for each phase individually. Thus, an LPV mixed-sensitivity weighting scheme that respects the requirements of each phase at each end of the domain is feasible. Consider the profile ($\theta_r$) is generated as a slew in a given time-frame from the initial $\theta_0$ to the desired attitude $\theta_r(t\rightarrow \infty)$. As in Fig.~\ref{fig:rhoDefine} initially the error from attitude to desired final attitude is large. As the spacecraft slews and the attitude converges to pointing, the error decreases. Therefore, the chosen scheduling parameter is $\rho = |\theta - \theta_r(t\rightarrow \infty)|$ as it quantitavely represents the transition from acquisition slew to target pointing. The scheduling parameter is defined within a range $\rho_p < \rho <\rho_a$, where the bounds are chosen as reasonable limits with respect to the two modes. In terms of implementation, for any value of $\rho$ where $\rho < \rho_{p}$ or $\rho > \rho_{a}$ the controller will behave as an LTI controller. This is of particular use for fine pointing since the scheduling parameter is a quasi-state of the controller; LTI behaviour when $\rho < \rho_{p}$  guarantees the best stability for fine-pointing.


	\begin{figure}
		\begin{minipage}{0.39\linewidth}
			\centering
				\begin{tikzpicture}
	
\begin{axis}[%
	width=1.5in,
	height=1.1in,
	at={(0in,1.2in)},
	scale only axis,
	xmin=0,
	xmax=1000,
	ymin=1,
	ymax=5.5,
	ylabel style={font=\color{white!15!black}},
	ylabel={Magnitude},
	axis background/.style={fill=white},
	axis x line*=bottom,
	axis y line*=left,
	xlabel = {Time},
	title ={\textcolor{white}{|}},
	legend style = {draw = none},
	ymajorticks = false,
	yticklabel = \empty,
	xtick = {112.8},
	xticklabel = {$T$},
	]

\addplot [color=mycolor3, line width = 1pt]
table[row sep=crcr]{%
	5.5	4.71238898038473\\
	67.2	4.70820436648035\\
	84.4	4.69565052476742\\
	101.6	4.67472745524594\\
	118.8	4.64543515791581\\
	136	4.60777363277703\\
	153.2	4.56174287982958\\
	170.4	4.50734289907359\\
	187.6	4.44457369050906\\
	204.8	4.37343525413576\\
	222	4.29392758995391\\
	239.2	4.20605069796352\\
	256.4	4.10980457816447\\
	273.6	4.00518923055677\\
	290.8	3.8922046551404\\
	308	3.7708508519155\\
	325.2	3.64112782088193\\
	342.5	3.50224497598356\\
	427.7	2.80527981937485\\
	444.9	2.67626557619485\\
	462.1	2.55562056082351\\
	479.3	2.44334477326083\\
	496.5	2.3394382135067\\
	513.7	2.24390088156122\\
	530.9	2.15673277742428\\
	548.1	2.077933901096\\
	565.3	2.00750425257638\\
	582.5	1.9454438318653\\
	599.7	1.89175263896288\\
	616.9	1.84643067386901\\
	634.1	1.8094779365839\\
	651.3	1.78089442710723\\
	668.5	1.76068014543932\\
	685.7	1.74883509157996\\
	703	1.74532925199435\\
	1005.6	1.74532925199435\\
};

\addlegendentry{$|\theta_r|$}

\addplot [color=mycolor4, line width = 1pt]
table[row sep=crcr]{%
	5.5	5.05486677646161\\
	65.7	5.0506828980898\\
	81.4	5.03813126297428\\
	97.1	5.01721187111502\\
	112.8	4.98792472251205\\
	128.5	4.95026981716546\\
	144.2	4.90424715507515\\
	159.9	4.84985673624112\\
	175.6	4.78709856066337\\
	191.3	4.715972628342\\
	207	4.63647893927691\\
	222.7	4.5486174934681\\
	238.4	4.45238829091568\\
	254.1	4.34779133161942\\
	269.8	4.23482661557955\\
	285.5	4.11349414279596\\
	301.2	3.98379391326876\\
	316.9	3.84572592699783\\
	332.6	3.69929018398307\\
	351.7	3.51184423171242\\
	426.2	2.78031338963456\\
	441.9	2.63779858757675\\
	457.6	2.50365154226267\\
	473.3	2.3778722536922\\
	489	2.26046072186546\\
	504.7	2.15141694678243\\
	520.4	2.05074092844302\\
	536.1	1.95843266684733\\
	551.8	1.87449216199536\\
	567.5	1.79891941388712\\
	583.2	1.7317144225226\\
	598.9	1.67287718790169\\
	614.6	1.6224077100245\\
	630.3	1.58030598889093\\
	646	1.54657202450119\\
	661.7	1.52120581685506\\
	677.4	1.50420736595265\\
	693.1	1.49557667179397\\
	712.1	1.49439510239324\\
};

\addlegendentry{$|\theta|$}

\draw [color=mycolor4, line width = 1pt] plot [smooth, tension=1] coordinates { (712.1, 1.4944) (1000,1.74533)};

\draw [color=black, <->, line width = 1pt] plot [] coordinates { (112.8,	4.98792472251205) (112.8,1.74532925199435)};
\node [circle,draw = none] at (250, 3) {$\rho_{t = T}$};
\draw [color=mycolor1, dashed, line width = 1pt] plot [] coordinates { (0,1.74532925199435) (1000,1.74532925199435)};

\end{axis}	
	
\end{tikzpicture}
		\captionof{figure}{Scheduling parameter definition}
		\label{fig:rhoDefine}
		\end{minipage}
\begin{minipage}{0.6\linewidth}
		\centering
	\begin{tikzpicture}

\begin{axis}[%
	width=1.1in,
	height=1in,
	at={(0in,0in)},
	scale only axis,
	separate axis lines,
	xmin=0,
	xmax=2.5,
	xtick = {0,1,2,3},
	xticklabels = {{},{$\omega_{e,p}$},{$\omega_{e,a}$},{}}, 
	ymin=-100,
	ymax=15,
	ytick = {-80, -50, 0},
	yticklabels ={$\epsilon_p$,$\epsilon_a$,$0$},
	ytick pos=left,
	xtick pos = bottom,
	ylabel={Magnitude (dB)},
	xlabel={Frequency},
	axis background/.style={fill=white},
	title={$W_e^{-1}$},
	]
	
	\addplot [color=black, dashed, forget plot, line width = 1pt]
	table[row sep=crcr]{%
		2 -100\\
		2  0 \\
	};
	
	\addplot [color=black, dashed, forget plot, line width = 1pt]
	table[row sep=crcr]{%
		1 -100\\
		1 0 \\
	};

		\addplot [color=black, dashed, forget plot, line width = 1pt]
	table[row sep=crcr]{%
		0 0\\
		2 0 \\
	};

\addplot [color=mycolor1, forget plot, line width = 1pt]
table[row sep=crcr]{%
 0  -80\\
0.5 -80 \\
1  0 \\
1.0375 6 \\
3 6 \\
};
\label{line:egptWeight},

\addplot [color=mycolor2, forget plot, line width = 1pt]
table[row sep=crcr]{%
	0  -50\\
	1.6875 -50 \\
	2  0 \\
	2.0375 6 \\
	3 6 \\
};
\label{line:egacqWeight},
	
\end{axis}

\begin{axis}[%
	width=1.1in,
	height=1in,
	at={(1.5in,0in)},
	scale only axis,
	separate axis lines,
	xmin=0.5,
	xmax=3.5,
	xtick = {1.75},
	xticklabels={$\omega_{u}$},
	ymin=-70,
	ymax= 40,
	ytick = {0,30},
	yticklabels ={$R_{eu,a}^{-1}$,$R_{eu,p}^{-1}$},
	ytick pos=left,
	xtick pos = bottom,
	xlabel={Frequency},
	axis background/.style={fill=white},
	title={$W_u^{-1}R_{eu}^{-1}$},
	legend style={legend cell align=left, align=left, draw=white!15!black}
	]
	
	\addplot [color=black, dashed, forget plot, line width = 1pt]
	table[row sep=crcr]{%
		1.75 -100\\
		1.75  30 \\
	};

		\addplot [color=mycolor1, forget plot, line width = 1pt]
	table[row sep=crcr]{%
		0 30\\
		1.75  30 \\
		2.5 -30 \\
		3.5 -30 \\
	};
	
	\addplot [color=mycolor2, forget plot, line width = 1pt]
	table[row sep=crcr]{%
		0 0 \\
		1.75 0 \\
		2.5 -60 \\
		3.5 -60 \\
	};

\end{axis}
\end{tikzpicture}
	\captionof{figure}{Frequency domain weights at $\rho_{p}$ (\ref{line:egptWeight}) and $\rho_{a}$ (\ref{line:egacqWeight})}
	\label{fig:LPVWeights}
\end{minipage}

\end{figure}


Consider first the pointing mode, the task of the controller is to maintain the attitude of the spacecraft with a steady-state error less than a given value $\epsilon_p$. Hence, as depicted in Fig.~\ref{fig:LPVWeights}, the shape of $W_e$ ensures the sensitivity function $S$ is pushed below $\epsilon_p$ at low frequency. Given that the satellite has integral dynamics, there will in reality be no steady-state error. Therefore, the main driver in the pointing performance is the disturbance rejection; the parameter $\epsilon_p R_{du}R_{eu}^{-1}$ describes the steady-state error response to a disturbance input. The tracking bandwidth $\omega_{e,p}$, also impacts the disturbance rejection performance so it must be higher than the frequency range of expected low frequency disturbance. The weight $W_u$ defines the shape of the actuator response to reference ($KS$) and disturbance signals ($-KSP$). Hence it is a first order weighting function with a high pass shape, enforcing that the actuator response rolls off at a given frequency $\omega_{u,p}$. This frequency reflects the available bandwidth in the system and ensures the controller rolls off to prevent the excitation of flexible modes or unmodelled high frequency dynamics. To maximise the tracking and disturbance rejection capabilities, $\omega_{e,p}$ can be further increased so long as the sensitivity peak $|S_{max}|$ remains low and the frequency separation between $\omega_{e,p}$ and $\omega_u$ is enough to provide sufficient phase margin (PM).

In comparison, the task of the acquisition mode is to track a slew guidance profile. So that the controller design can be posed as an LPV problem, the tunable design parameters for the acquisition mode are the same as for the pointing mode. The main differences in the design are that the steady-state error $\epsilon_a$ in this phase does not need to be as low as for the pointing phase. Additionally, there is more emphasis on tracking bandwidth $\omega_{e,a}$, the controller response should be faster as the spacecraft is performing a manoeuvre. The acquisition mode typically uses thrusters to produce torque, which have a considerably higher torque capacity than RW. However, given that this paper considers a fine-pointing scenario, the allowable error will be orders of magnitude higher for the acquisition mode than the pointing mode. Thus the main influence on $R_{eu}$ are the pointing requirements, hence $R_{eu,p} < R_{eu,a}$. Similarly $R_{du,a}$ is greater to account for larger torque disturbances relative to the actuator capabilities during the manoeuvre.

The focus of this work is on the controller algorithm and smoothing the command and error signal. Therefore, implementing realistic actuator dynamics are outside of the scope of this paper. However, it is important to note that there exist torque allocation algorithms that would enable the combined use of thrusters and reaction wheels (RW). An example of such an algorithm is \cite{groette2023constrained}.

Now the two design points can be reformulated as an LPV induced $\mathcal{L}_2$-norm synthesis problem. The plant dynamics are considered unchanging during the manoeuvre, so the plant $P$ is linear time-invariant (LTI). The two mixed-sensitivity weighting schemes are interpolated across the domain with respect to a chosen function shape (e.g. linear, quadratic, hyperbolic). Thus, the problem is formulated as a grid-based LPV mixed-sensitivity controller synthesis with parameter-varying performance weights. The parameter-varying generalised plant $G_\rho$ takes the form in (\ref{eqn:lpvsystem}). Then the controller, as in (\ref{eqn:K}), is found by solving the optimisation problem in (\ref{eqn:minimisation}). 
In the present paper, the controller design is formulated as a quadratic function of $\rho$. A quadratic rather than a linear function was chosen so that the rate of change in the controller begins low while the spacecraft is slewing, and then increases when the spacecraft gets closer to pointing - when tracking errors are lower. This was the simplest, low order, rational domain shape that produced good results. Linear, inverse and hyperbolic funcions were also investigated. For the linear function, performance was not satisfactory due to the constant rate of change. For the inverse and hyperbolic functions, the performance showed no improvement but the synthesis time was significantly longer and the problem definition became overly complex. An example for the definition of $R_{eu}$ is provided below in (\ref{eqn:quadDomain}). The equation ensures the end points of the function correspond to the pointing and acquisition design points.

\begin{subequations}\label{eqn:quadDomain}
	\begin{align}
		R_{eu}(\rho) = & R_{eu,a} + \alpha_{eu}(\rho - \rho_{a})^2 \\
		\alpha_{eu} = & \frac{R_{eu,p}-R_{eu,a}}{(\rho_p - \rho_{a})^2} 
	\end{align}
\end{subequations}


\section{APPLICATION: Target Acquisition and Fine Pointing of a Flexible Satellite}

\begin{figure}[ht]
	\centering
\includegraphics[width = 0.5\linewidth]{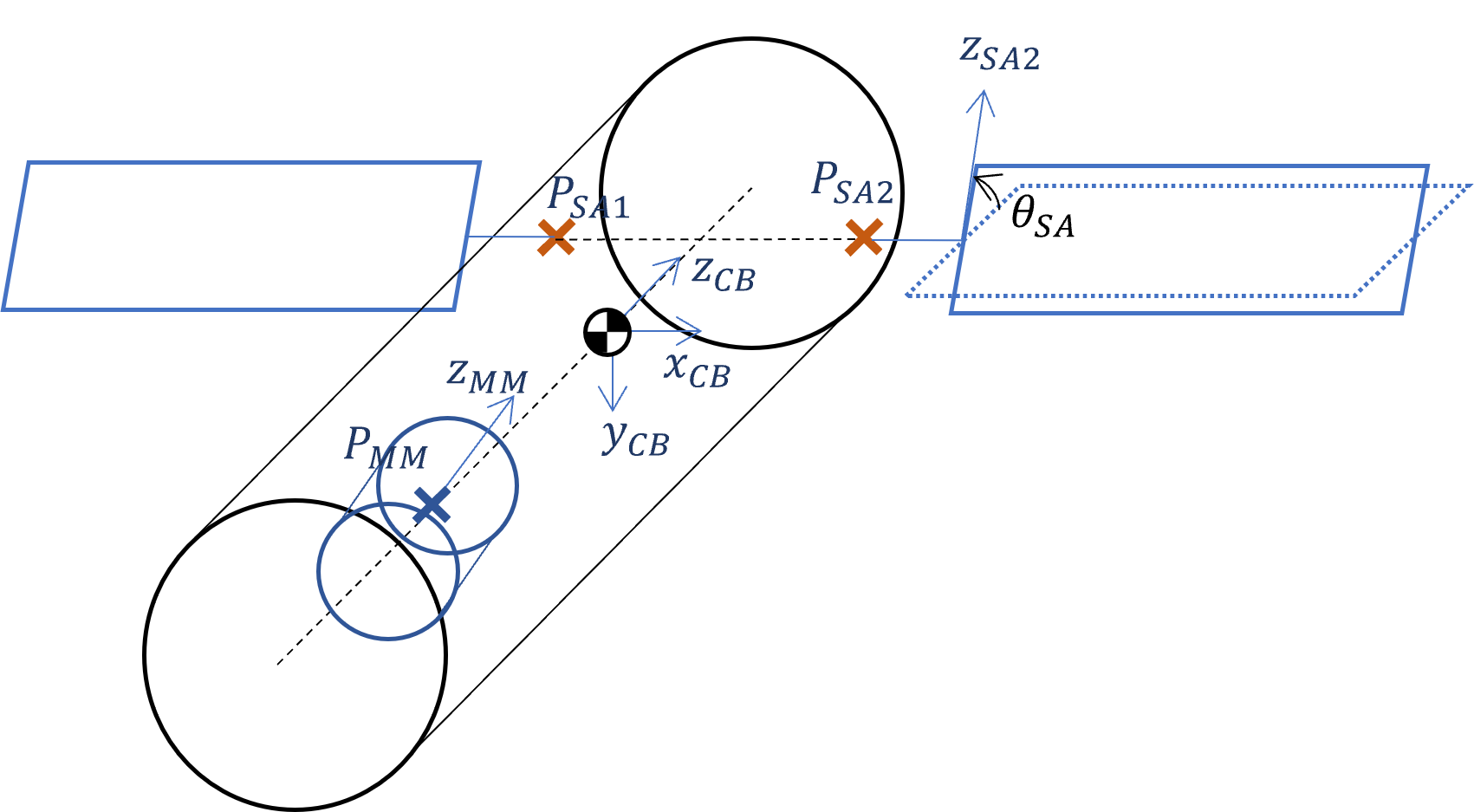}
\caption{Diagram of satellite model characteristics.}
\label{fig:satModel}
\end{figure}

The proposed LPV controller design is demonstrated on a simulation of a flexible satellite. The satellite in question is a large observation satellite (similar to Chandra X-Ray NASA, or Athena ESA), weighing $6000$~kg. It was modelled using the Satellite Dynamics Toolbox \cite{SdtWebpage,alazard2020satellite,sanfedino2018finite,chebbi2017linear,sanfedino2019experimental}. As depicted in Fig. \ref{fig:satModel}, the satellite has two flexible solar arrays which are able to rotate in unison. Also, it has a moveable observation module (weighing an additional $1000$~kg) positioned at the opposite end to the solar arrays. The spacecraft manoeuvres by generating torques about the body frame in order to align its $z$-axis in the body frame $z_{\text{CB}}$ to the target. The attitude controller must satisfy certain requirements in the frequency and time domain, representative of typical industry requirements. First, the dynamics of the flexible appendages should not be excited. Then, for additional robustness, the open-loop system must have gain and phase margins of at least $6\,$~dB and $35^\circ$, respectively. Finally, the system must have a modulus margin of at least $0.5$, which is equivalent to the magnitude of the output sensitivity function remaining below $6\,$~dB. 

In terms of time domain performance, the tracking and pointing performance is measured by the absolute performance error (APE). In other words, it is the error between the reference attitude $\theta_r$ and true attitude $\theta$. In addition, the actuator commands must remain within provided actuator bounds $\tau_{max}$. This bound corresponds to thruster torque authority during the acquisition phase. Then actuator commands must converge to below the reaction wheel saturation limits during pointing phase. Time domain requirements and limitations are summarised in Table \ref{tab:timeReq}.

\begin{table}
	\centering
	\caption{System requirements and limitations.}
	\label{tab:timeReq}
	\begin{tabular}{r | c | c}
		& \textbf{Acquisition} & \textbf{Pointing} \\ \hline \hline
		APE $[x,y,z]$ [rad] & $[0.026,0.026,0.026]$ & $[7.3\times 10^{-6}, 7.3\times 10^{-6},0.175]$ \\
		$\tau_{max}$ $[x,y,z]$ [Nm] & $[50,25,20]$ & $[0.27,0.14,0.15]$
	\end{tabular}
\end{table}

The central body of the satellite is modelled as a cylinder and is described by the linear Newton-Euler equations (\ref{eqn:centralBody}) with mass $m_{\text{CB}}$ and inertia $J_{\text{CB}}$.

\begin{equation} \label{eqn:centralBody}
\left(\begin{array}{c}
	
	\Sigma f_{\text{ext}}\\
	\Sigma \tau_{\text{ext}}
\end{array}\right) = 
\left[\begin{array}{c c}
	m_{\text{CB}} I_{3\times3} & 0_{3\times3}\\
	0_{3\times3} &  J_{\text{CB}}
\end{array}\right]
\left(\begin{array}{c}
	\ddot{r}\\
	\ddot{\theta}
\end{array}\right)
\end{equation}

The sum of external forces $f_{\text{ext}}$ and torques $\tau_{\text{ext}}$ acting on the spacecraft result in a translational and rotational acceleration ($\ddot{r}$ and $\ddot{\theta}$ respectively). The solar arrays (SA) are each modelled as a cantilever beam connected to the central body at a hinge point $P$.

\begin{equation} \label{eqn:flexAppendage}
\begin{array}{l}
	\left(\begin{array}{c}
		f^P\\
		\tau^P
	\end{array}\right) = 
	\left[\begin{array}{c c}
		m_{\text{SA}} I_{3\times3} & 0_{3\times3}\\
		0_{3\times3} &  J_{\text{SA}}\
	\end{array}\right]
	\left(\begin{array}{c}
		\ddot{r}_{\text{SA}}\\\
		\ddot{\theta}_{\text{SA}}\
	\end{array}\right) + L_P^T\ddot{\eta} \\
	-L_P \left(\begin{array}{c}
		\ddot{r}_{\text{SA}}\\\
		\ddot{\theta}_{\text{SA}}\
	\end{array}\right) = \ddot{\eta} + diag\{2\zeta_i \omega_i\}^k_{i = 1}\dot{\eta} + diag\{ \omega_i^2\}^k_{i = 1}\eta
\end{array}
\end{equation}
$L_{P}$ describes the modal contributions of the solar arrays. Each second-order mode (denoted by subscript $i$) has damping $\zeta_i$ and natural frequency $\omega_i$. The solar arrays in the model include the first and second bending modes and a torsional mode with natural frequencies $[5.6,35.0,19.3]$~rad/s respectively. The moveable observation module is modelled as a smaller rigid cylinder connected to the central body at point $P$ by rotational transformations. As with the rotation of the solar arrays, these rotational transformations are included as uncertainties in the design, this ensures the controller achieves performance for all solar array and observation module positions. All uncertainties in the model are summarised in Table \ref{tab:uncertainties}.

\begin{table}[t]
	\caption{System parameters modelled with uncertainty.}
	\centering
	\label{tab:uncertainties}
	\begin{tabular}{l l| c c}
		& Parameter & Nominal & Range \\ \hline \hline
		Central body (CB)
		& Mass $m_{\text{CB}}$ [kg] & $6000$ & $\pm 5\%$  \\
		Distance from origin $O$ & $OG_x$ [m] & $0$ & $\pm 0.15$ \\
		to CoG position $G$ & $OG_y$ [m] & $0$ & $\pm 0.15$ \\
		 & $OG_z$ [m] & $7.5$ & $\pm 5\%$ \\  \hline
		Measurement Module (MM)
		& Mass $m_{\text{MM}}$ [kg] & $1000$ & $\pm 5\%$ \\
		& Rotation-$x_{\text{MM}}$ [$^\circ$] & $0$ & $\pm 0.5$\\
		& Rotation-$y_{\text{MM}}$ [$^\circ$] & $0$ & $\pm 5$\\
		& Rotation-$z_{\text{MM}}$ [$^\circ$] & $0$ & $\pm 0.5$\\  \hline
		Solar Arrays (SA)
		& Rotation  $\theta_{\text{SA}}$ [$^\circ$] & $0$ & $\pm 180$ \\ \hline
	\end{tabular}
\end{table}

Some approximate disturbance and sensor noise models are introduced in order to test the robustness of the controller against external factors. The expected disturbance, namely disturbance torque from the space environment, is approximated as white noise passed through a low pass filter. The navigation algorithm is not a focus of this paper, so it is assumed that the satellite navigates with sensor fusion. This results in a white noise introduced to the output channel, the magnitude of which corresponds to the typical temporal error in a star tracker.
The considered slew manoeuvre GUI profile is defined by a constant acceleration towards the target attitude followed by a constant deceleration such that the final attitude is reached with no residual angular rate. Various slew sizes were considered, with a change in attitude ranging from $20^\circ$ to $180^\circ$. All slews were generated such that the GUI profile reaches the final attitude in $700$s. Given that the spacecraft model has higher torque capacity about the x-axis, this is the preferred direction of rotation for performing a larger slew.

%

%

\subsection{Controller Design}

\begin{figure}
\centering
\input{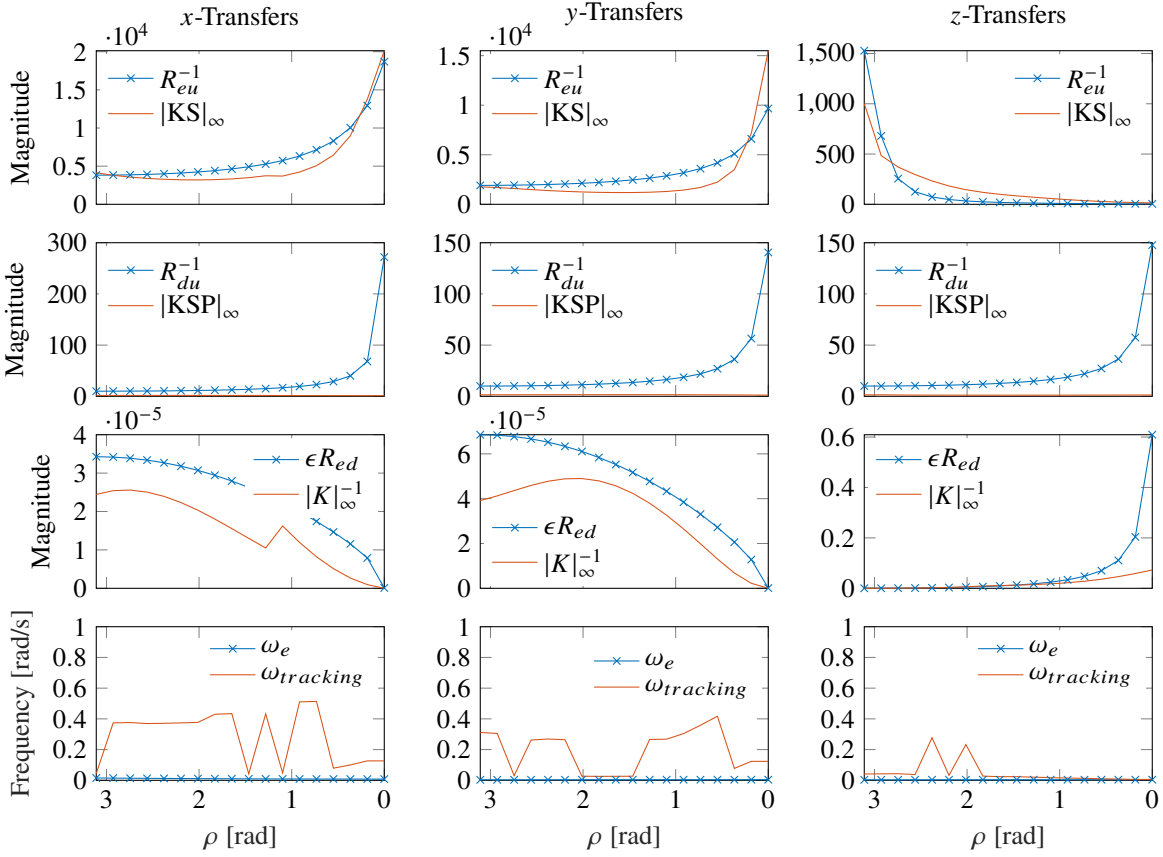}
\caption{Tuning parameters over the domain and corresponding closed-loop properties.}
\label{fig:tuningParams}
\end{figure}

\begin{figure}[t]
\centering
\input{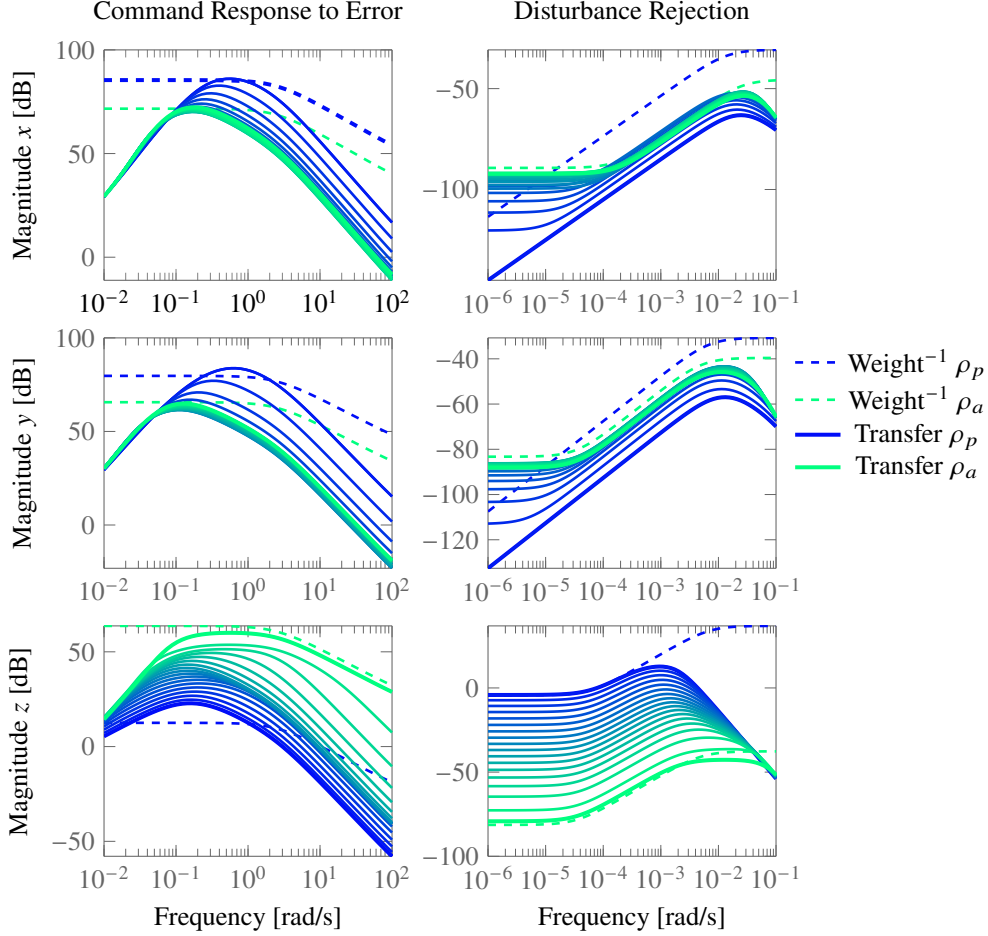}
\caption{Frequency response of closed-loop transfer functions across the gridded domain $\boldsymbol{KS}$ (left) and $\boldsymbol{-KSP}$ (right) for each axis.}
\label{fig:4blocks}
\end{figure}

The controller synthesis used the nominal rigid body dynamics of the satellite derived in (\ref{eqn:centralBody}) and (\ref{eqn:flexAppendage}), removing the flexible modes of the appendages. This minimised the number of states in the full-order controller and the design allows enough frequency separation that it does not excite the flexible modes in the full order plant. All weights in the design are diagonal, corresponding to the three axes of the system. All tuned parameters are interpolated between the end points using the quadratic function (\ref{eqn:quadDomain}). To encompass a full range of slew manoeuvres and ensure the controller only converges to LTI once fine-pointing is sustained: $\rho_p = 0.01^\circ$ ($1.75\times10^{-4}$~rad) and $\rho_{a} = 180^\circ$ ($\pi$~rad).

The shapes of $W_e$ and $W_u$ are designed as in Fig.~\ref{fig:LPVWeights}. Across the domain, $W_e$ is set to $-6$~dB at high frequency to limit the peak of the sensitivity function to $2$. The magnitude of $W_u$ increases by $40$~dB after the crossover frequency $\omega_u = 2.5$~rad/s which is chosen as half of the frequency of the first flexible mode. This ensures sufficient roll-off in the controller so that the commands are lower within the frequency range of the solar array flexible modes. Since the required roll-off is unchanging, $W_u$ is not parameter-varying. The steady-state tracking error $\epsilon$ changes from $\epsilon_a = [0.013,0.013,0.013]$ to $\epsilon_p = [7.3\times 10^{-6},7.3\times 10^{-6},0.0175]$. For pointing $\epsilon_p$ is equal to the APE requirement (see Table \ref{tab:timeReq}), whereas for acquisition $\epsilon_a$ is half the corresponding APE requirement to improve the tracking performance for this mode. It is very important that the RW do not saturate. For this reason, the ratio $R_{eu}$ was tuned such that, for pointing, the controller will saturate the RW only if the error reaches two times $\text{APE}_p$; $R_{eu,p} = (2\times \text{APE}_p)/(\tau_{max,p}) = [5.3\times 10^{-5},10\times10^{-5},0.24]$. On the other hand, for acquisition, fast tracking is the priority, so $R_{eu}$ corresponds to an actuator saturating at half $\text{APE}_a$. In other words $R_{eu,a} = (0.5\times \text{APE}_a)/(\tau_{max,a}) = [0.26,0.52,0.65]\times 10 ^{-3}$. Similarly, the ratio $R_{du}$ is tuned for pointing so that the RW will not saturate for the expected disturbance $R_{du,p} = [3.7,7.1,6.8]\times 10^{-3}$. The ratio is low, meaning the actautors can be more responsive to small disturbances. For acquisition, this ratio is increased to $R_{du,a} = [0.1,0.1,0.1]$. The ratio $R_{ed} = R_{eu}R_{du}^{-1}$ results from the other tuned parameters, but it is still important to check that the result is realistic as this directly relates to the disturbance rejection capabilities of the system. As described in Section \ref{sec:LPVTrans}, $\epsilon R_{ed}$ corresponds to the amplification of low frequency disturbance in the attitude error. The resulting disturbance rejection is $\epsilon_p R_{ed,p} = [1.06\times 10 ^{-7},1.06\times 10^{-7},0.61]$. For the x- and y-axes this is equivalent to a steady-state error of approximately $1\%$ of the APE requirement resulting from an input disturbance of $1$~Nm. For the z-axis, the disturbance rejection is relaxed as the requirements are much less stringent but the level of disturbance is the same. Above $10^{-4}$~rad/s it is possible that disturbances will be amplified, but given the relative size between expected disturbance and allowable error, and the frequency range this occurs, it is deemed acceptable. For acquisition, the amplification is of similar magnitude across the axes $\epsilon_a R_{ed,a} = [3.4,6.9,8.6]\times 10^{-5}$. Note that the expected disturbance is higher during acquisition so its important that all three axes have sufficiently small values. The final parameter to tune was the tracking bandwidth $\omega_e$. This was iteratively pushed up to maximise the achievable tracking and disturbance rejection performance of each mode while keeping enough frequency separation to meet robustness margin requirements. The final tunings are $\omega_{e,p} = [8,4,4]\times 10^{-3}$~rad/s and $\omega_{e,a} = [1.5,0.3,0.3]\times 10^{-2}$~rad/s. It is not imperative that the tuning occurs in this order, however it is important that all tuning parameters and their corresponding physical constraints on the system are carefully considered.

For synthesis, the LPV mixed-sensitivity weights were interpolated across the domain on a grid of 18 points such that the synthesis problem was a finite selection of LMIs. The resulting parameter tunings are plotted in Fig.~\ref{fig:tuningParams} together with the resulting controller and closed-loop characteristics they influence. It is clear from Fig.~\ref{fig:tuningParams} that the design drivers are the disturbance rejection limit ($\epsilon R_{ed}$) and the relationship between actuator commands and tracking error ($R_{eu}^{-1}$). The controller was synthesised using \texttt{LPVTools} \cite{hjartarson2015lpvtools}. The trajectory of $\rho$ in the parameter dependent storage functions was chosen as $p_0 + p_1\rho^2 + p_2\rho^4$, where $p$ are coefficients solved in the optimisation. The rate-bounds of $\dot{\rho}$ were chosen corresponding to the largest expected slew rate, $[-8.7\, 8.7]\times 10^{-3}$ rad, which comes from the GUI profile definition.

Fig.~\ref{fig:4blocks} shows the resulting closed-loop transfer functions with the LPV controller compared to the weighting scheme at the two end points, as described by the four-block-problem. It is clear from the plot that the synthesised LPV controller closely follows the dynamics imposed by the design process. It also demonstrates that the synthesis was able to achieve higher frequency tracking and disturbance rejection bandwidths than tuned.

\subsection{Evaluation}
\begin{figure}[bt!] 
	\centering\input{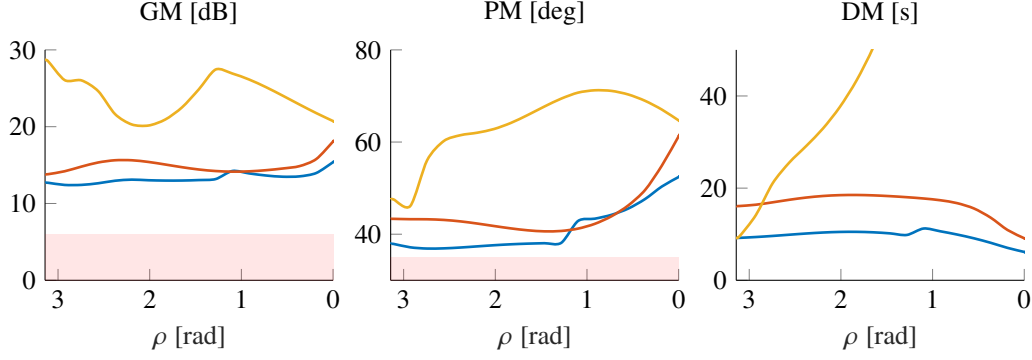}
	\caption{Worst-case margins over all uncertainties across the parameter domain. Margins of transfers to the output x-axis (\ref{line:xaxis}), y-axis (\ref{line:yaxis}) and z-axis (\ref{line:zaxis}) attitude.}
	\label{fig:marginsLPV}
\end{figure}

The evaluation and simulations were conducted with the full-order plant, including flexible modes. The worst stability margins of the LPV design were calculated by taking a random sample of 100 LTI plants over all uncertainties in Table \ref{tab:uncertainties}. The results are plotted in Fig.~\ref{fig:marginsLPV} with respect to the scheduling parameter. All margin requirements are met. Generally, the gain and phase margins (GM and PM) show an upward trend as the system converges to pointing, which is desirable as the system is more robust when performance is more critical to the mission. On the other hand, the delay margin (DM) reduces for the x- and y-axes to $6.3$s and $9.4$s respectively.

\begin{table}[t]
	\caption{Monte-Carlo distribution parameters. Normal distributions $\mathcal{N}(\mu,\sigma)$ with mean $\mu$ and standard deviation $\sigma$. Uniform distributions $\mathcal{U}(\mu + \text{range})$}
	\centering
	\label{tab:MCinputDist}
	\begin{tabular}{l l| c}
		& Parameter & Distribution \\ \hline \hline
		\multirow{2}{*}{Central Body (CB)} 
		& Mass $m_{\text{CB}}$ [kg] & $\mathcal{N}(6000,1.7\%)$ \\
		& $OG_x$ [m] & $\mathcal{N}(0,0.05)$ \\
		& $OG_y$ [m] & $\mathcal{N}(0,0.05)$ \\
		& $OG_z$ [m] & $\mathcal{N}(7.5,0.13)$ \\  \hline
		\multirow{1}{*}{Measurement Module (MM)}
		& Mass $m_{\text{MM}}$ [kg] & $\mathcal{N}(1000,1.7\%)$\\
		& Rotation-$x_{\text{MM}}$ [$^\circ$] & $\mathcal{U}(0 \pm 0.5)$\\
		& Rotation-$y_{\text{MM}}$ [$^\circ$] & $\mathcal{U}(0 \pm 5)$\\
		& Rotation-$z_{\text{MM}}$ [$^\circ$] & $\mathcal{U}(0 \pm 0.5)$\\  \hline
		Solar Arrays (SA)
		& Rotation  $\theta_{\text{SA}}$ [$^\circ$] & $\mathcal{U}(0 \pm 180)$ \\ \hline
		Initial error   & $|\theta_{r,x}(0)|$ [$^\circ$] & $\mathcal{U}(20 + 160)$ \\ 
		& $|\theta_{r,y}(0)|$ [$^\circ$] & $\mathcal{U}(20 + 70)$ \\ \hline
	\end{tabular}
\end{table}

\begin{figure}[ht!]
\centering
\begin{tikzpicture}
	
	\begin{axis}[%
		width=2in,
		height=1in,
		at={(3in,0in)},
		scale only axis,
		ymin = 0,
		ymax = 1000,
		xmin = 1300,
		xmax = 1700,
		ylabel={Frequency [\%]},
		ytick = {0,500,1000},
		yticklabels = {{$0$},{$10$},{$20$}},
		scaled ticks=false,
		xlabel={Convergence time [s]},
		axis background/.style={fill=white},
		axis x line*=bottom,
		axis y line*=left,
		axis on top,
		title ={$|\theta_e|$ converged below APE},
		]
		\addplot[line width=0.75pt, draw=mycolor1, fill=mycolor1!50,ybar, bar width = 10] coordinates {
			(1335,2)	(1345,55)	(1355,80)	(1365,111)	(1375,135)	(1385,156)
			(1395,164)	(1405,159)	(1415,212)	(1425,231)	(1435,285)	(1445,282)
			(1455,310)	(1465,302)	(1475,364)	(1485,397)	(1495,418)	(1505,499)
			(1515,533)	(1525,508)	(1535,582)	(1545,659)	(1555,715)	(1565,736)	
			(1575,772)	(1585,669)	(1595,382)	(1605,171)	(1615,77)	(1625,25)
			(1635,8)	(1645,0)	(1655,1)
		};
		
		
	\end{axis}
	
		\begin{axis}[%
		width=2in,
		height=1in,
		at={(0in,0in)},
		scale only axis,
		ymin = 0,
		ymax = 4000,
		xmin = 700,
		xmax = 900,
		ylabel={Frequency [\%]},
		ytick = {0,1250,2500,3750,5000},
		yticklabels = {{$0$},{$25$},{$50$},{$75$},{$100$}},
		scaled ticks=false,
		xlabel={Convergence time [s]},
		axis background/.style={fill=white},
		axis x line*=bottom,
		axis y line*=left,
		axis on top,
		title ={$\tau$ converged within RW limits (all axes)},
		]
		
				\addplot[line width=0.75pt, draw=mycolor1, fill=mycolor1!50,ybar, bar width = 10] coordinates {
						(715,3807)
						(815,6)
						(825,83)
						(835,227)
						(845,417)
						(855,416)
						(865,44)
					};
		
	\end{axis}
	
\end{tikzpicture}
\caption{Distribution of time taken to converge to fine pointing.}
\label{fig:MCdist}
\end{figure}
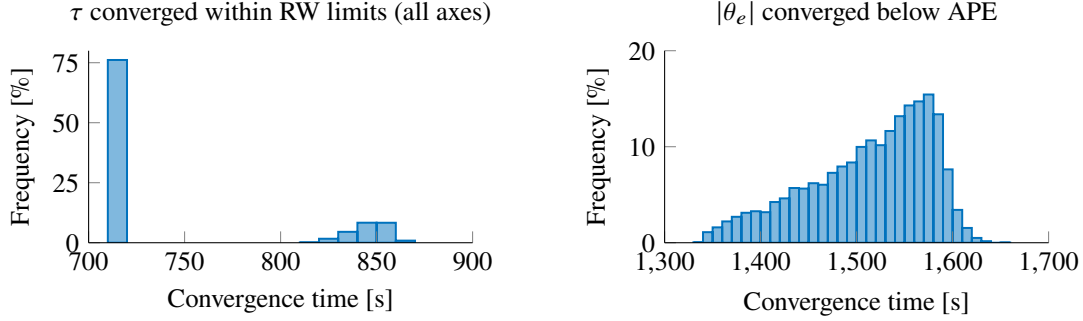

\begin{figure}[ht!]
	\centering
	\input{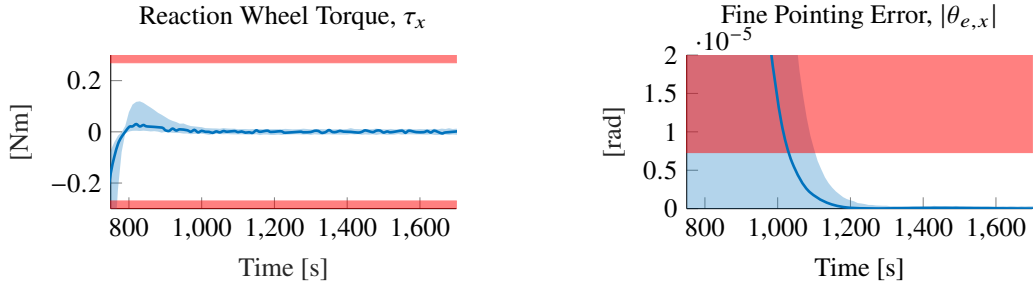}
	\caption{Time response of the LPV controller across the MC campaign (\ref{fill:MCresponse}), with a nominal case (\ref{line:MCNomResponse}). Showing fine-pointing is achieved smoothly with respect to requirements (\ref{fill:Req}).}
	\label{fig:MCtimeCloseUp}
\end{figure}

\begin{figure}[ht!]
\centering
	\input{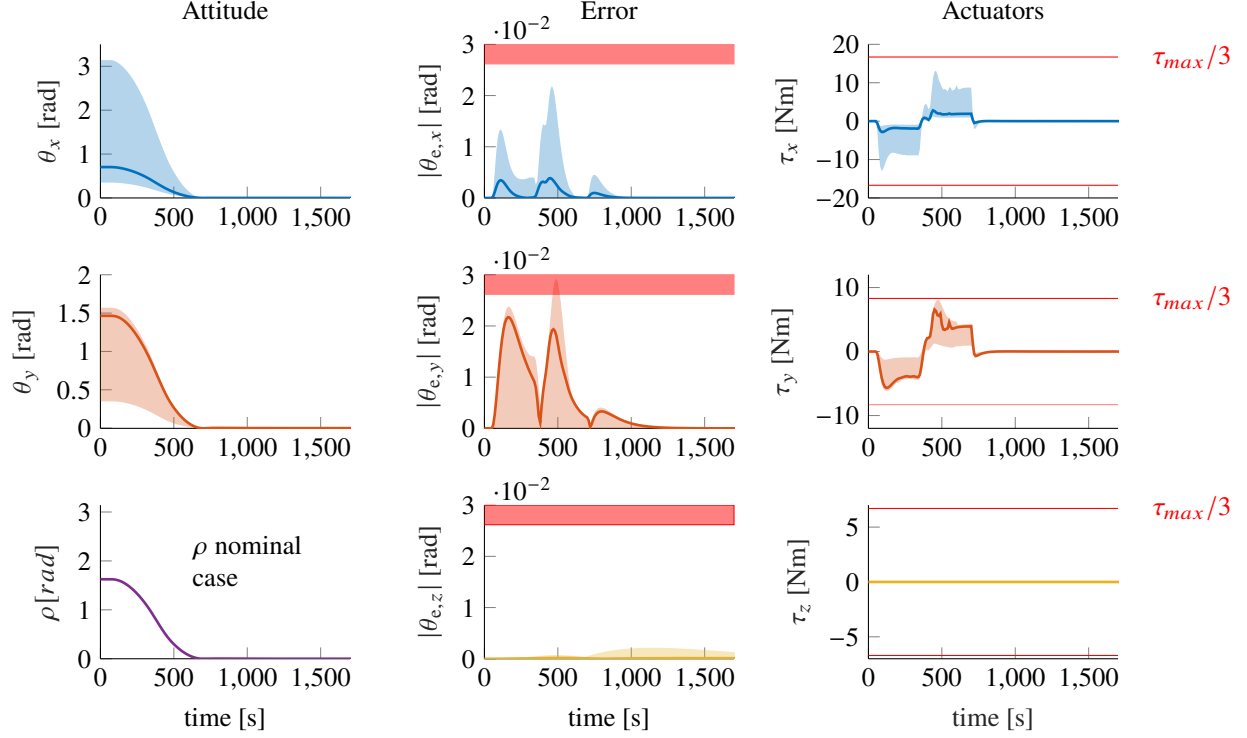}
\caption{Time response of the LPV controller across the MC campaign (\ref{line:x},\ref{line:y},\ref{line:z}), with the nominal case. Showing slew performance with respect to aquisition requirements (\ref{fill:Req}).}
\label{fig:MCtime}
\end{figure}

The proposed LPV controller was assessed through a Monte-Carlo campaign with 10,000 samples. The flexible satellite model was implemented in \texttt{Simulink} and simulated with a distribution of initial attitude errors and plant uncertainties, as summarised in Table \ref{tab:MCinputDist}. The modelled low frequency input disturbance and white noise had unique noise generation seeds for each simulation. As the goal of the design was to achieve fine pointing of a target as fast as possible after completing a slew, the measure of performance was the time to converge to within the pointing requirement bound $|\theta_e|\leq \text{APE}$ given in Table \ref{tab:timeReq}. The results in Fig.~\ref{fig:MCdist} show that, for every case, the torque command $\tau$ converges to below the RW limits before the attitude error $|\theta_e|$ reaches fine-pointing requirements. This reflects the current industry practice of switching to RW before converging to fine-pointing. The majority of cases converge to torque commands below RW command limits within $750$~s, less than a minute after the slew is complete. The remaining cases exhibit a larger overshoot in the $y$-axis command, and so converge in the next oscillation. Fine-pointing convergence shows a smooth distribution with the fastest convergence time at $\approx1335$~s; which is $\approx 635$~s (under $11$ minutes) after the GUI profile converges to steady-state. The slowest case takes $\approx 955$~s (under $16$ minutes) to converge to fine-pointing after the GUI profile is complete. Fig.~\ref{fig:MCtimeCloseUp} further shows that as the attitude and torque commands converge, the signals are smooth and there are no undesireable jumps in the torque command, as would occur in a discrete switch implementation. Similar behaviour is seen for the $y$-axis, and the $z$-axis states remain withing pointing requirements throughout the simulations. Fig.~\ref{fig:MCtime} shows that, with the exception of $\theta_{e,y}$, all limitations and requirements during the slew are met over the Monte-Carlo campaign. This requirement is only not met briefly, for cases where the initial $\theta_x$ is towards the maximum limit and $\theta_y$ is at its minimum limit.


\section{CONCLUSION}

An LPV controller design scheduled with attitude error was proposed for the combined control and smooth mode transition between an acquisition and pointing mode of a spacecraft mission. The LPV approach leads to a single controller with guaranteed performance bounds across the domain with respect to the tuning. Moreover, the dynamics of the controller are known over the transition period and stability margins can easily be computed over the domain. The design used a mixed-sensitivity control scheme with parameter-varying weights that are derived from the performance and robustness requirements and system limitations. The design was implemented on a spacecraft with a moveable measurement module, flexible solar arrays, sensor noise, and external torque disturbances. The use-case was acquiring and achieving fine-pointing of a target within given requirements. The results of a Monte-Carlo campaign confirmed that the LPV controller converged to pointing smoothly with a range of initial attitudes and system states (with respect to uncertainties) and was able to successfully bring the actuator commands within the reaction wheel saturation limits.



%
%
	\section*{Acknowledgments}
	This work was supported by the ESA OSIP co-funded research program under the contract 4000140113/22/NL/GLC/my, entitled \textit{AI-Enhanced Real-Time Identification and LFT-Based Control for Space Systems}. It was also partially supported by the European Union under Grant No. 101153910 entitled \textit{Recommissioning and Deorbiting using Cubesat Swarms with Electro Spray Thrusters}. Views and opinions expressed are however those of the authors only and do not necessarily reflect those of the European Union. Neither the European Union nor the granting authority can be held resposnible for them.
	
	
	\bibliography{literatur.bib}
	
\end{document}